\documentclass[conference]{IEEEtran}

\ifCLASSINFOpdf
\else
\fi
\usepackage{amsmath}
\usepackage{amsfonts}
\usepackage[caption=false,font=small,labelfont=sf,textfont=sf]{subfig}
\usepackage[linesnumbered,ruled,vlined]{algorithm2e}
\usepackage{amssymb}
\usepackage{subcaption}

\usepackage{xcolor}
\usepackage{flushend}
\usepackage{comment}
\usepackage{graphicx}
\usepackage{algorithm2e}
\RestyleAlgo{ruled}
\usepackage{algpseudocode}
\usepackage{amsmath}
\usepackage{textcomp}
\usepackage{epsfig}
\usepackage{multicol}
\usepackage[justification=raggedright,singlelinecheck=false]{caption}
\usepackage[colorlinks=true, linkcolor=blue, citecolor=blue, urlcolor=blue]{hyperref}
\usepackage{algpseudocode}
\usepackage{tikz}
\usepackage{geometry}
\usepackage{pgfplotstable}
    \pgfplotsset{
        compat=1.9,
        compat/bar nodes=1.8,
}

\usepackage{acronym}
\acrodef{ce}[CE]{Channel Estimation}
\acrodef{nr}[NR]{New Radio}
\acrodef{graphnet}[GraphNet]{Graph Neural Estimation Network}
\acrodef{mimo}[MIMO]{Multiple Input Multiple Output}
\acrodef{siso}[SISO]{Single Input Single Output}
\acrodef{ls}[LS]{Least Squares}
\acrodef{mmse}[MMSE]{Minimum Mean Square Error}
\acrodef{mse}[MSE]{Mean Square Error}
\acrodef{dl}[DL]{Deep Learning}
\acrodef{cnn}[CNN]{Convolutional Neural Network}
\acrodef{dmrs}[DM-RS]{Demodulation Reference Signals}
\acrodef{ofdm}[OFDM]{Orthogonal Frequency Division Multiplexing}
\acrodef{mmimo}[mMIMO]{Massive MIMO}
\acrodef{gnn}[GNN]{Graph Neural Network}
\acrodef{re}[RE]{Resource Element}
\acrodef{qam}[QAM]{Quadrature Amplitude Modulation}
\acrodef{awgn}[AWGN]{Additive White Gaussian Noise}
\acrodef{prb}[PRB]{Physical Resource Block}
\acrodef{ut}[UT]{User Terminal}
\acrodef{sr}[SR]{Super Resolution}
\acrodef{ir}[IR]{Image Restoration}
\acrodef{srcnn}[SRCNN]{CNN-based SR}
\acrodef{dncnn}[DnCNN]{CNN-based Denoising}
\acrodef{relu}[ReLU]{Rectified Linear Unit}
\acrodef{nn}[NN]{Neural Network}
\acrodef{dnn}[DNN]{Deep Neural Network}
\acrodef{gcn}[GCN]{Graph Convolutional Network}
\acrodef{cdf}[CDF]{Cumulative Distribution Function}
\acrodef{snr}[SNR]{Signal to Noise Ratio}
\acrodef{bs}[BS]{Base Station}
\acrodef{tdl}[TDL]{Tapped Delay Line}
\acrodef{li}[LI]{Linear Interpolation}
\acrodef{mlp}[MLP]{Multi Layer Perceptron}
\acrodef{pdsch}[PDSCH]{Physical Downlink Shared Channel}
\acrodef{tdl}[TDL]{Tapped Delay Line}
\acrodef{bler}[BLER]{Block Error Rate}
\acrodef{cir}[CIR]{Channel Impulse Response}
\acrodef{csi}[CSI]{Channel State Information}
\begin{document}
\title{Transformer Actor-Critic for Efficient Freshness-Aware Resource Allocation\\
}

\author{
\IEEEauthorblockN{Maryam Ansarifard\IEEEauthorrefmark{1}, Mohit K. Sharma\IEEEauthorrefmark{2}, Kishor C. Joshi\IEEEauthorrefmark{1}, George Exarchakos\IEEEauthorrefmark{1}}

\IEEEauthorblockA{\IEEEauthorrefmark{1}Department of Electrical Engineering, Eindhoven University of Technology, Eindhoven, The Netherlands\\
Emails: \{m.ansarifard, k.c.joshi, G.Exarchakos\}@tue.nl}
\IEEEauthorblockA{\IEEEauthorrefmark{2}Directed Energy Research Center, Technology Innovation Institute, Abu Dhabi, UAE\\
Email: mohit.sharma@tii.ae}
}
\maketitle
\begin{abstract}
Emerging applications such as autonomous driving and industrial automation demand ultra-reliable and low-latency communication (URLLC), where maintaining fresh and timely information is critical. A key performance metric in such systems is the age of information (AoI). This paper addresses AoI minimization in a multi-user uplink wireless network using non-orthogonal multiple access (NOMA), where users offload tasks to a base station. The system must handle user heterogeneity in task sizes, AoI thresholds, and penalty sensitivities, while adhering to NOMA constraints on user scheduling. We propose a deep reinforcement learning (DRL) framework based on proximal policy optimization (PPO), enhanced with a Transformer encoder. The attention mechanism allows the agent to focus on critical user states and capture inter-user dependencies, improving policy performance and scalability. Extensive simulations show that our method reduces average AoI compared to baselines. We also analyze the evolution of attention weights during training and observe that the model progressively learns to prioritize high-importance users. Attention maps reveal meaningful structure: early-stage policies exhibit uniform attention, while later stages show focused patterns aligned with user priority and NOMA constraints. These results highlight the promise of attention-driven DRL for intelligent, priority-aware resource allocation in next-generation wireless systems.
\end{abstract}


\section{Introduction}
In the era of ubiquitous connectivity and intelligent edge devices, computation offloading has emerged as a pivotal mechanism for enabling scalable and efficient processing in wireless networks. Computation offloading involves transferring processing tasks from resource-constrained devices, such smartphones, Internet-of-Things (IoT) sensors and autonomous systems. This paradigm significantly enhances system performance by reducing computational latency, conserving local energy resources, and supporting the execution of sophisticated applications beyond the capability of edge devices alone. 

Ultra-reliable low-latency communication (URLLC) in next-generation wireless networks imposes stringent latency and reliability requirements on computation offloading, especially for mission- and safety-critical applications. Prior work \cite{van2022urllc} studies latency minimization in DT-enabled wireless edge networks using URLLC links by jointly optimizing transmission and computation resources. Computation offloading is essential in latency-sensitive applications such as AR \cite{morin2022toward}, real-time analytics, autonomous driving, healthcare, and industrial automation, where information freshness is critical. However, wireless offloading faces challenges from stochastic channels, limited spectrum, and rapidly varying network conditions, which can lead to delayed or outdated task execution.

To mitigate the limitations of spectrum scarcity and improve spectral efficiency in multi-user environments, non-orthogonal multiple access (NOMA) has emerged as a promising multiple access technique for wireless communication systems. Unlike traditional orthogonal schemes, NOMA enables simultaneous transmission to multiple users over the same time and frequency resources by leveraging power domain multiplexing and successive interference cancellation (SIC) \cite{maraqa2020survey}. This makes NOMA particularly attractive for computation offloading scenarios in dense networks, where efficient user scheduling and resource sharing are essential to maintain low latency and high reliability.

For URLLC applications, the age of information (AoI) is a key metric for quantifying information freshness, capturing the time since the most recently received update was generated \cite{kosta2017age}. Unlike delay or throughput, AoI provides a more accurate measure of timeliness but is difficult to minimize in offloading environments due to dynamic resource allocation and uncertainty. Prior work has studied AoI optimization in MEC systems using analytical approaches such as queuing theory and stochastic geometry \cite{dong2024mean, tang2023age}. However, these classical methods often struggle with scalability and adaptability in dynamic, large-scale settings.

Deep reinforcement learning (DRL) has emerged as an effective approach for AoI-aware optimization in wireless systems, enabling adaptive decision-making under dynamic network conditions. Recent works apply DRL to diverse scenarios, including UAV-enabled edge networks \cite{ansarifard2024ai}, RIS-assisted vehicular systems \cite{10663259}, and time-sensitive IoT offloading \cite{jiang2023age}, using algorithms such as SAC, DDPG, and D3QN. Other studies address AoI-aware resource management in V2V and autonomous platooning via DRL and MARL frameworks \cite{chen2020age, parvini2023aoi}, demonstrating the strong capability of DRL in dynamic and high-mobility environments.

While these DRL-based frameworks show promising results, many still face limitations in capturing long-range dependencies and processing high-dimensional state spaces, often encountered in multi-user, real-time systems. In this context, attention mechanisms, originally introduced for sequence modeling in natural language processing \cite{vaswani2017attention}, provide a compelling enhancement. Attention allows models to dynamically focus on the most salient parts of the input, enhancing both interpretability and learning efficiency. When integrated into DRL architectures \cite{niu2021review, li2020deep}, attention mechanisms enable agents to prioritize critical updates, model structured interdependencies among users, and adaptively process state information in complex networked environments.

To further enhance scalability and user coordination, we propose a novel deep reinforcement learning framework that leverages Transformer-based attention mechanisms to optimize scheduling and power control in a multi-user NOMA system for AoI minimization. The practical realization of high-performance uplink NOMA, specifically the robustness and complexity of the Successive Interference Cancellation (SIC) receiver at the Base Station, remains a significant ongoing challenge in the physical layer domain. This paper focuses on the development of an optimal resource allocation policy assuming an ideal NOMA physical layer, representing the theoretical performance ceiling achievable when robust SIC is fully realized in next-generation wireless systems. The development of the DRL policy serves to prove the potential gains under these conditions, guiding future system design. The key contributions of the work in this paper are as follows:
\begin{itemize}
    \item We formulate an AoI-aware resource allocation problem for NOMA uplink as a constrained MDP with heterogeneous user task sizes, AoI thresholds, and penalty levels.
    \item We propose a scalable, priority-aware PPO-based scheduling policy augmented with a Transformer encoder to capture inter-user dependencies.
    \item We analyze the learned attention weights and show that the policy progressively focuses on high-priority users, revealing structured and interpretable attention patterns aligned with NOMA constraints.
    \item We show that attention improves both interpretability and AoI performance by exploiting asymmetric inter-user relationships in multi-user wireless systems.
\end{itemize}



\section{System Model}
We consider multiple users denoted as $\mathcal{U}=\{1, \dots U\}$ offloading their computation tasks denoted as $\mathcal{M}=\{1,\dots, M\}$ to a base station (BS), where $U$ and $M$ are total number of users and tasks, respectively. Time is considered to be slotted and normalized to the slot duration (i.e., the slot duration is taken as $1$). Uplink transmissions to the BS follow NOMA communication, where the decoding order at the BS is based on users’ received signal strengths, and SIC is applied to separate overlapping signals. Specifically, users with stronger signals are decoded after canceling the interference from weaker ones. The frequency band consists of a set of subcarriers (SCs) $\mathcal{N} = \{1, \dots, N\}$, where $N$ is the total number of subcarriers. Let $K_{u}^{n}(t)$ be an indicator showing whether subcarrier $n$ is assigned to user $u$ at time $t$; $K_{u}^{n}(t)=1$ if assigned, and $0$ otherwise. Note that at each time, a user can only transmit its task through one subcarrier, i.e. $C 1: \:\sum_{n =1}^{N}K_{u}^{n}(t) \leq 1, \forall u \in \mathcal{U}$. Based on practical NOMA considerations, each channel can be assigned to at most $2$ users, i.e. $C 2: \:\sum_{u =1 }^{U}K_{u}^{n}(t) \leq 2, \forall n \in \mathcal{N}$. 

For simplicity, we consider a block Rayleigh fading wireless channel model where the channel gains between users and channels follow an exponential distribution with unit mean. The channel gains denoted as $g_{u}^{n}(t)$ are assumed to remain constant within each time slot but vary independently across time slots. To model the decoding order in uplink NOMA, let $\pi_n(t)$ denote the ordered set of users sharing subcarrier $n$ at time $t$, arranged in ascending order of their channel gains (i.e., from weakest to strongest). That is, if user $u$ is decoded before user $v$, then $u < v$ in $\pi_n(t)$. In this order, users with stronger channels decode and cancel the interference from users with weaker channels using SIC. Accordingly, the data rate of user $u$ on subcarrier $n$ at time $t$ is given by \cite{ali2016dynamic} as:
\begin{align} \label{eq:sic_rate}
    R_u^n(t) = B \log_2\left(1 + \frac{p_u^n(t)g_{u}^{n}(t)}{\sigma + \sum\limits_{\substack{i \in \pi_n(t) \\ i > u}} K_{i}^{n}(t) p_i^n(t)g_{i}^{n}(t)}\right),
\end{align} 
where $B$ and $p_{u}^{n}(t)$, are the bandwidth allocated per subcarrier and transmit power allocated to user $u$ on SC $n$, respectively. $\sum\limits_{i > u} p_i^n(t)$ represents the residual intra-channel interference from users whose signals are decoded after user $u$, and $\sigma$ is a small constant representing the noise power.\\
\indent Considering maximum power of $p^{\text{max}}$, allocated power to each user should satisfy
\begin{equation}
	C 3: \:\sum_{n \in \mathcal{N}}K_{u}^{n}(t) p_{u}^{n}(t) \leq p^{\text{max}}, \forall u \in \mathcal{U}. 
\end{equation}

At each time \( t \) when a user has access to the channel to transmit its task, a portion of the task is transmitted. This fraction corresponds to the user's data rate. The remaining portion of the task on the user's side, denoted as \( L_{u,m}(t) \), can be expressed as:
\begin{align}
    &L_{u,m}(t) = L_{u,m}(t - 1) -R_u^n(t), \\\nonumber & L_{u, m}(0) = L_{u, m},
\end{align}
where $ L_{u, m}$ is the task size which we assume to be different for each user.\\
\indent The AoI evolves in discrete time steps, following a modified sawtooth pattern distinct from classical AoI systems. At each time step, the AoI increases by one unit for all users, representing the aging of information, calculated as:
\begin{align}
    a_u(t+1) = \begin{cases}
a_u(t) - a_{u,\text{reset}}(t) + 1, & \text{if } L_{u,m}(t) \leq 0, \\
a_u(t) + 1, & \text{otherwise}.
\end{cases}
\end{align}

When a task is completed ($L_{u,m}(t)\leq 0$), the AoI undergoes a partial reset rather than returning to zero. The reduction depends on the accumulated reset value $a_{u,\text{reset}}(t)$, which records the AoI at the last successful transmission and preserves transmission history in the AoI evolution. For example, if a user has AoI 3 at task generation and completes transmission after 4 time slots (AoI 7), the AoI resets to $7-3+1=5$, reflecting the generation-to-delivery delay, and the reset value is updated accordingly.
\subsection{Problem Formulation}
We aim to minimize both the average AoI across all users and the total weighted count of users whose AoI exceeds a predefined threshold, by allocating channel and transmission power of the assigned users at each time slot. The average AoI is normalized by the maximum allowed AoI $a_{u}^{\max}$ to bring both parts of the objective function to same scale.
\begin{subequations} \label{opt}
\begin{align}
\min_{\mathbf{K},\, \mathbf{P}_{\mathcal{U}}} \quad 
& \sum_{t=0}^{T}{\Big(\frac{1}{U \cdot a_{u}^{\max}} \sum_{u=1}^{U} a_u(t) + \lambda \sum_{u=1}^{U} w_u \cdot \mathbf{1}_{(a_u(t) > \tau_u)}\Big)}\label{opt_obj} \\
\text{s.t.} \quad 
& {C1-C3}, \notag \\
\label{opt_constraints}
\end{align}
\end{subequations}
where $\textbf{K}(t)=\{K_{u}^{n}(t), u \in \mathcal{U}, n \in \mathcal{N}\}$ and $\textbf{P}_{\mathcal{U}}(t)=\{p_{u}^{n}(t), u \in \mathcal{U}, n \in \mathcal{N}\}$ denote the channel and transmission power
allocation of all users, respectively. In the above, $\tau_u$ denotes the predefined AoI threshold for user $u$, which, in general, is different for each user. $w_u$ denotes the weight assigned to user $u$; a higher weight implies a greater penalty for surpassing the AoI threshold. To optimize tail performance, $w_u$ should be set such that $w_u \gg 1$, ensuring that any violation of $\tau_u$ incurs a substantial penalty and thus significantly degrades performance. In consistency with practical transceivers, we use discrete transmit power levels, with each user assigned to a channel selecting one of these levels. The decision vector $\mathbf{K}(t)$ is selected by the DRL policy at the start of each time slot, allowing users to be dynamically re-assigned to different subcarriers in subsequent slots based on the updated system state (e.g., AoI and channel conditions). For a given time slot, the problem in \eqref{opt} is a combinatorial optimization with discrete channel and power variables, further complicated by decision coupling across time slots. This makes analytical solutions intractable, motivating the use of an RL-based approach.
\section{Solution}
In this section, we model the optimization problem as an MDP and propose an attention-enhanced PPO algorithm, which is a policy-gradient method well-suited for high-dimensional discrete action spaces and continuous state environments.
\subsection{Reinforcement Learning-Based Solution Using PPO}

To solve the optimization problem using RL, we model the system as an MDP, defined by the tuple \((\mathcal{S}, \mathcal{A}, \mathcal{P}, \mathcal{R}, \gamma)\), where:

\begin{itemize}
    \item \textbf{State} (\(\mathcal{S}\)): At time \(t\), the state \(s_t\) includes the AoI vector of all users $\mathbf{a}(t) = [a_1(t), \dots, a_U(t)]$, task in users' sides $\mathbf{L}(t) = [L_{1,m}(t), \dots, L_{U,m}(t)], \forall m \in \mathcal{M}$, AoI reset values $\mathbf{a}_{\text{reset}}(t) = [a_{1,\text{reset}}(t), \dots, a_{U, \text{reset}}(t)]$, current channel conditions $\mathbf{g}(t) = [g_1^{n}(t), \dots, g_U^{n}(t)], \forall n \in \mathcal{N}$, described as follows:
    \begin{align} \label{state}
	\mathbf{s}_{t} = \{\mathbf{a}(t), \mathbf{L}(t), \mathbf{a}_{\text{reset}}(t), \mathbf{g}(t)\}.
    \end{align}
     The state is defined by the instantaneous channel conditions $g(t)$. Due to DRL inference delay $\tau_{inf}$, policies in practice act on slightly outdated information $g(t-\tau_{inf})$. For learning and benchmarking, our simulations assume perfect CSI $g(t)$, a standard simplification in theoretical DRL resource allocation studies that informs deployment considerations.
    \item \textbf{Action} (\(\mathcal{A}\)): The action $\mathbf{a}(t) \in \mathcal{A}$ comprises the joint channel and power allocation decisions (taking values in a discrete set.), specifically $\{\mathbf{K}(t), \mathbf{P}(t)\}$, where $K_{u}^{n}(t)$ and $P_{u}^{n}(t)$ are determined instantaneously at the start of time slot $t$.
    
    \item \textbf{Transition} (\(\mathcal{P}\)): The environment transitions based on the current state and action, influenced by stochastic dynamics governing varying channel conditions and packet arrivals.
    
    \item \textbf{Reward} (\(\mathcal{R}\)): In a time slot, the reward is designed as the negative of the  instantaneous value of the objective function in (\ref{opt_obj}).
    \begin{align}
       \mathbf{r}_t = -\left(\frac{1}{U \cdot a_{u}^{\max}} \sum_{u=1}^{U} a_u(t) + \lambda \sum_{u=1}^{U} w_u \cdot \mathbf{1}_{(a_u(t) > \tau_u)}\right). 
    \end{align}
     Therefore, maximizing the reward corresponds to minimizing AoI and penalty terms:
    \item \textbf{Discount Factor} (\(\gamma\)): A factor \(0 < \gamma < 1\) is used to balance immediate versus future rewards.
\end{itemize}

PPO is well suited for our problem because it handles discrete action spaces (user–channel assignments and quantized power levels) and ensures stable training through a clipped surrogate objective, avoiding the instability of vanilla policy gradients. The soft constraints in (5a) are incorporated into the reward as penalties, while the hard constraints in (5b) are strictly enforced. To guarantee feasible actions, we use a two-stage mechanism: (1) Feasibility pre-filtering (C1, C2), where the set of NOMA-feasible user–subcarrier pairings is pre-computed, ensuring at most one subcarrier per user and at most two users per subcarrier; and (2) Action masking and clipping (C3), where infeasible actions are masked before softmax and power outputs are clipped to satisfy maximum power constraints. 
\subsubsection*{PPO Training Procedure}

The agent interacts with the environment across multiple episodes. At each timestep, it observes the current state $\mathbf{s}_t$ and selects an action $\mathbf{a}_t$. Upon executing the action, the agent receives a reward designed to penalize both AoI increase and threshold violations. To assess the quality of the taken action, the \textit{critic network} estimates the value function \(V(\mathbf{s}_t)\). The \textit{advantage function} \(\hat{A}_t\), which guides the policy update, is estimated using generalized advantage estimation (GAE) \cite{schulman2015high}:
\begin{align}
    &\hat{A}_t = \sum_{l=0}^{\infty} (\gamma \lambda_{s})^l \delta_{t+l}, \quad  \delta_t = \mathbf{r}_t + \gamma V(\mathbf{s}_{t+1}) - V(\mathbf{s}_t),
\end{align}
where $V(\mathbf{s}_t)$ is the value function approximating expected returns, and $\lambda_s$ is a smoothing parameter. PPO then updates the policy by maximizing the clipped surrogate objective:
\begin{align}
    L^{\text{CLIP}}(\theta) = \mathbb{E}_t \left[ \min \left( v_t(\theta) \hat{A}_t, \, \text{clip}(v_t(\theta), 1 - \epsilon, 1 + \epsilon) \hat{A}_t \right) \right],
\end{align}
where \(v_t(\theta) = \frac{\pi_\theta(\mathbf{a}_t | \mathbf{s}_t)}{\pi_{\theta_{\text{old}}}(\mathbf{a}_t | \mathbf{s}_t)}\) and $\epsilon$ are the probability ratio between the new and old policies, and a small constant that controls the clipping range, respectively. The overall PPO objective combines three components: the clipped policy loss, the value function loss, and an entropy bonus to encourage exploration. The total loss function is expressed as:
\begin{align}
    L^{\text{PPO}} = L^{\text{CLIP}} - c_1L^{\text{VF}} + c_2S[\pi_\theta],
\end{align}
where $L^{\text{VF}} = (V_\theta(\mathbf{s}_t) - \mathbf{r}t)^2$ denotes the value function loss, $S[\pi\theta]$ is the policy entropy that promotes exploration, and $c_1$ and $c_2$ weight the value loss and entropy bonus, respectively. This actor–critic formulation, together with PPO’s clipped objective, enables stable learning of a policy that minimizes average AoI and the weighted penalties for violating user-specific AoI thresholds.
\subsection{Attention-Based PPO with Transformer Actor-Critic}
To further enhance the decision-making capability of the agent, we propose an attention-based PPO variant where the actor and critic networks are built upon Transformer encoders. The key idea is to capture inter-user dependencies through self-attention mechanisms, enabling the agent to make more informed and globally-aware scheduling and resource allocation decisions. In the standard PPO setup, the actor and critic process user features independently or with limited interaction. However, in dynamic wireless environments, users' AoI evolution, interference, and resource contention are inherently interdependent, specially in NOMA systems. Therefore, we adopt a Transformer-based design where each user’s feature vector is contextualized with respect to other users via multi-head self-attention. This is achieved through the self-attention mechanism, which operates on the triplet of queries (Q), keys (K), and values (V). Each user's state $\mathbf{s}_i$ is first embedded into a latent representation. These embeddings are then linearly projected into three distinct representations: \textit{queries} $\mathbf{q}_i$, \textit{keys} $\mathbf{k}_i$, and \textit{values} $ \mathbf{v}_i$, through learned linear projections:
\begin{align}
   \mathbf{q}_i = \mathbf{W}^Q \mathbf{x}_i, \quad
\mathbf{k}_i = \mathbf{W}^K \mathbf{x}_i, \quad
\mathbf{v}_i = \mathbf{W}^V \mathbf{x}_i, 
\end{align}
where $\mathbf{W}^Q, \mathbf{W}^K, \mathbf{W}^V \in \mathbb{R}^{d_{\text{model}} \times d_{\text{in}}}$ are learnable weight matrices, $d_{\text{model}}$, and $d_{\text{in}}$ are the dimensionality of the Transformer model and input state vector, respectively.
The attention weights between user $i$ and user $j$ are computed using the scaled dot-product attention:
\begin{align}
    \alpha_{ij} = \frac{
    \exp\left( \frac{\mathbf{q}_i^\top \mathbf{k}_j}{\sqrt{d_{\text{model}}}} \right)
}{
    \sum_{j'=1}^{N} \exp\left( \frac{\mathbf{q}_i^\top \mathbf{k}_{j'}}{\sqrt{d_{\text{model}}}} \right)
}.
\end{align}

These weights determine how much attention user $i$ should pay to user$ j$’s information. The output representation for each user $i$ is then computed as the weighted sum of value vectors:
\begin{align}
    \mathbf{z}_i = \sum_{j=1}^{N} \alpha_{ij} \mathbf{v}_j.
\end{align}

This resulting context vector $\mathbf{z}_i$ captures the relational importance of user $i$’s state in the context of all other users. These enriched representations are then passed through feedforward layers and ultimately fed into the PPO policy network to decide which users to schedule and which channel to assign. Using this mechanism, the agent strongly prioritizes users whose AoI is approaching critical thresholds or who have more urgent computation tasks, without relying on manually designed prioritization schemes. Moreover, the learned attention weights \( \alpha_{ij} \) offer interpretability, revealing the agent’s internal logic for prioritizing updates over time.
\begin{figure}[t]
    \centering
    \includegraphics[width=0.5\textwidth]{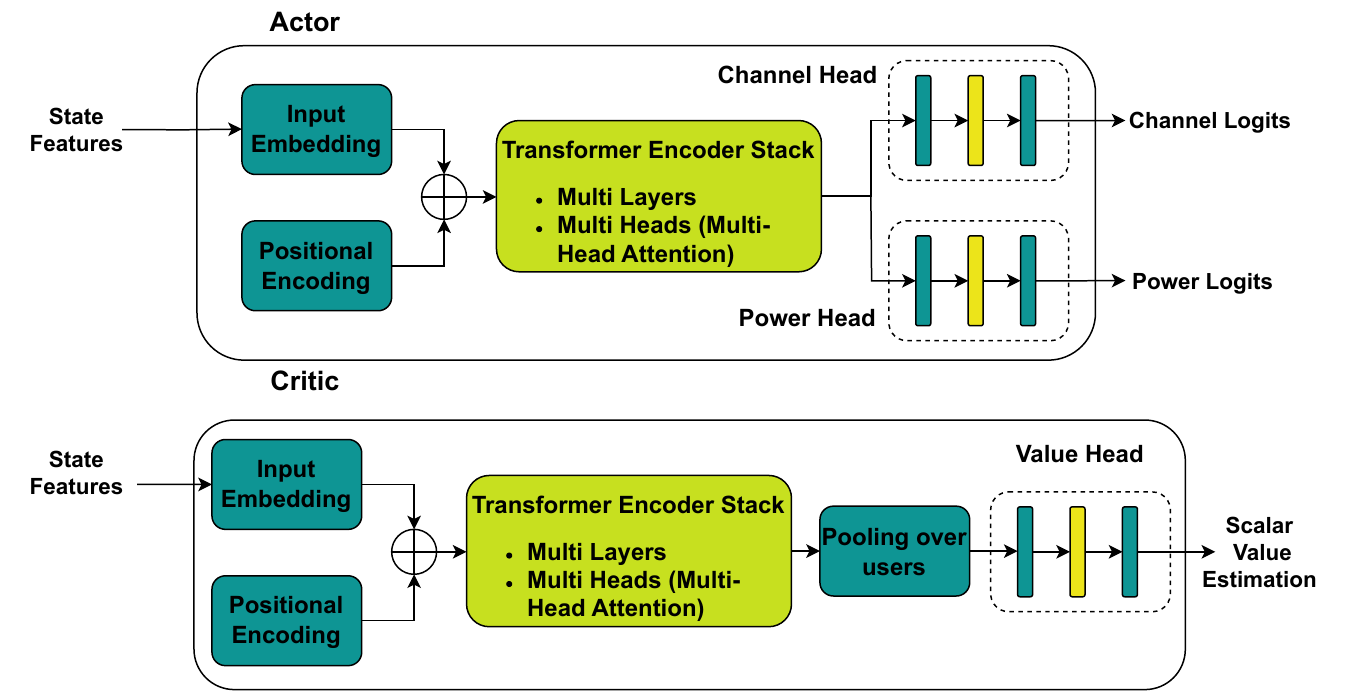}
   \caption{Transformer-based PPO Actor-Critic architecture. The encoder captures user-to-user dependencies before outputting action logits and value estimations. Each of the Channel, Power, and Value Heads comprises a different type of layer in sequence: linear, ReLU, and linear, respectively.}
    \label{architecture}
\end{figure}
\subsubsection*{Architecture Overview}

The overall architecture is illustrated in Fig. \ref{architecture}. The input to the model is a batch of per-user state features which include the AoI vector of all users, task in users' sides, AoI reset values, and current channel conditions. These are embedded into a high-dimensional space, enriched with positional encodings, and passed through multiple Transformer layers. Unlike traditional architectures that treat user states independently, self-attention layers in Transformers allow the agent to capture complex inter-user effects, such as competition for channels and correlated AoI spikes, by weighting each user's contribution dynamically.
\subsubsection*{Actor Network}
This module receives a batch of user-wise input features of shape \([B, U, d]\), where \(B\) is the batch size, \(U\) is the number of users, and \(d\) is the input dimension. Each feature vector is first projected to a shared embedding space and enhanced with learned positional encodings. A stack of Transformer encoder layers then computes context-aware embeddings for each user, capturing inter-user relationships. These embeddings are fed into separate output heads to generate per-user logits for both channel selection and power level allocation.
\subsubsection*{Critic Network}
This module follows a similar architecture but outputs a scalar value estimation. After obtaining user embeddings via the Transformer encoder, the critic aggregates them to form a global state representation. This pooled representation is passed through a feedforward head to predict the state value. 

This Transformer-based actor-critic framework allows the policy to reason about users jointly, rather than in isolation. The attention mechanism enables dynamic weighting of user interactions, which is particularly beneficial for optimizing metrics like AoI where performance is coupled across users. Empirically, this leads to improved sample efficiency and better tail-performance in threshold-aware scheduling tasks.

\section{EXPERIMENTS AND NUMERICAL RESULTS}
We consider NOMA wireless communication system in which users share frequency channels to transmit data to a BS. In order to study the impact of inter-user interference on achievable data rates, we allow the BS to allocate up to two users per channel, creating scenarios with controlled interference. This setup is designed to evaluate whether an attention-based mechanism can effectively capture and model the dependencies between users, both across all users in the system and specifically among users sharing the same channel. To capture diverse quality-of-service requirements, each user is assigned a unique AoI violation threshold. These thresholds form an arithmetic sequence, beginning at $15$ time slots for User $1$ and increasing by $1$ time slot for each additional user. Similarly, users are allocated different computational task sizes, also arranged in an arithmetic sequence starting at $1$ Mbit for User $1$ and increasing by $0.25$ Mbits per user. To reflect varying user priority levels, violation penalty weights are assigned in a linearly decreasing manner. Specifically, these weights follow a descending arithmetic progression, starting at $40$ for User $1$ and decreasing by $2$ for each subsequent user. The simulation is conducted using the PyTorch library on an Nvidia L40S GPU. The simulation environment and algorithm configuration parameters used in our experiments are summarized in Table~\ref{tab:simulation_parameters}.
\begin{table}[h!]
\captionsetup{justification=centering,singlelinecheck=false}
\centering
\caption{Simulation Parameters}
\begin{tabular}{|l|c|}
\hline
\textbf{Parameter} & \textbf{Value} \\
\hline
Number of users & 20 \\
Number of subcarriers & 8 \\
Total system bandwidth & 1 MHz \\
Fading parameter & 1.0 \\
Maximum transmission power & 0.1 W \\
Maximum tasks per user & 3 \\
PPO buffer capacity & 16384 \\
Transformer model dimension & 256 \\
Number of transformer heads & 8 \\
Number of transformer layers & 3 \\
Learning rate (actor and critic) & $5 \times 10^{-5}$ \\
Epsilon ($\epsilon$) & 0.2 \\
Clipping parameter $c_1$ & 0.5 \\
Clipping parameter $c_2$ & 0.05 \\
Number of PPO update epochs & 4 \\
Batch size & 64 \\
GAE parameter ($\lambda_s$) & 0.97 \\
Discount factor ($\gamma$) & 0.99 \\
Number of episodes & 50000 \\
\hline
\end{tabular}
\label{tab:simulation_parameters}
\end{table}

\subsection{Benchmarking Transformer-based PPO}
\begin{figure}[!t]
    \centering
    \includegraphics[width=\columnwidth]{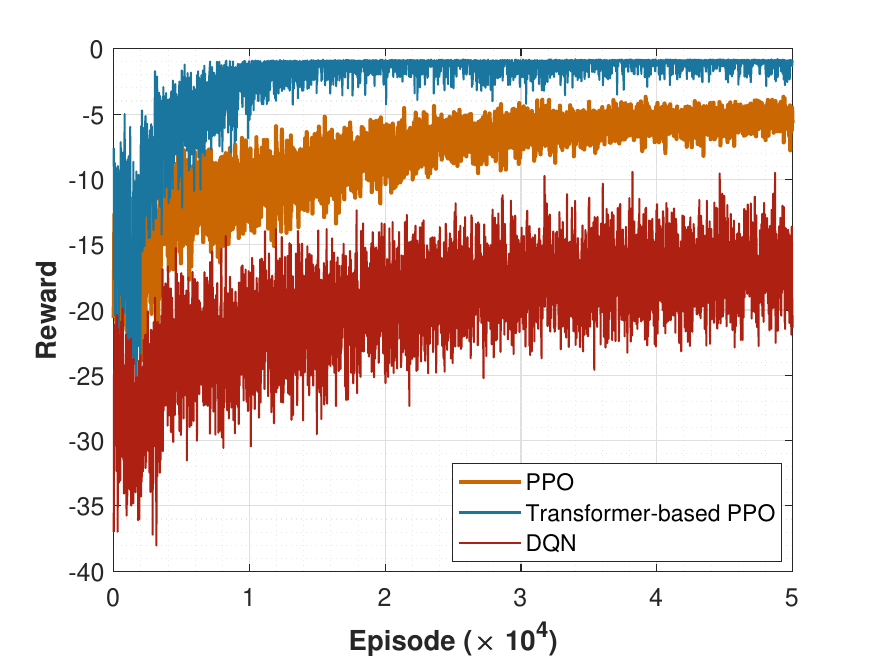}
    \caption{Training performance comparison of different RL approaches for multi-user NOMA resource allocation.}
    \label{reward}
\end{figure}

To evaluate the specific impact of the attention mechanism, we perform an ablation study by substituting the Transformer encoder layers in the PPO framework with a simple multi-layer perceptron (MLP), applied independently to each user's input. This modification removes any ability for the model to share information among users. Additionally, given the discrete nature of the action space, we include a comparison with the DQN algorithm. Fig.~\ref{reward} illustrates the training performance of three reinforcement learning algorithms, DQN, PPO with a MLP, and the proposed Transformer-based PPO in terms of average cumulative reward over $5 \times 10^4$ episodes.

The proposed Transformer-based PPO method demonstrates superior performance, achieving the highest reward among the three approaches. It not only converges faster but also stabilizes at a near-optimal reward level, showcasing its capability to model long-range dependencies and prioritize heterogeneous user states effectively. This improvement can be attributed to the attention mechanism, which allows the agent to focus on critical state features, such as task size, AoI threshold, and penalty weight, leading to more informed and efficient scheduling decisions. The standard PPO with MLP while outperforming DQN, still falls short of the transformer approach, highlighting that traditional fully-connected architectures struggle to capture the nuanced interactions between users competing for limited channel resources. While it benefits from the inherent stability of policy gradient methods, it lacks the structural ability to model inter-user relationships, resulting in suboptimal convergence. The DQN's poor performance suggests that value-based methods may be inherently less suitable for this multi-user coordination problem, potentially due to the large discrete action space (channel-power combinations for multiple users) and the need for exploration in a complex cooperative-competitive environment.

\subsection{Analysis of Attention Mechanism in Transformer-based PPO}
To study the learning dynamics of the Transformer-based PPO agent in a multi-user NOMA system, we analyze attention weights across training episodes. The attention mechanism operates over the user dimension, capturing inter-user dependencies for joint channel and power allocation. Each attention matrix is $20 \times 20$, with entry $(i, j)$ representing the attention user $i$ assigns to user $j$ during decision-making. 

\subsubsection{User Priority Structure}

The users in the system are heterogenous in terms of their task sizes, AoI thresholds, and penalty functions. Specifically, User~1 has the lowest task size, the strictest AoI threshold, and the highest penalty for exceeding it. User~2 has the second-lowest task size, the second-strictest threshold, and the second-highest penalty, and so on. User~20, conversely, has the largest task size, the most relaxed threshold, and the lowest penalty. This structured prioritization implies that the scheduling policy should pay more attention to high-priority users (e.g., User~1) in order to minimize the overall AoI penalty.
\begin{figure}[htbp]
    \centering

    \subfloat[Attention weights in the first episode]{%
        \includegraphics[width=\linewidth]{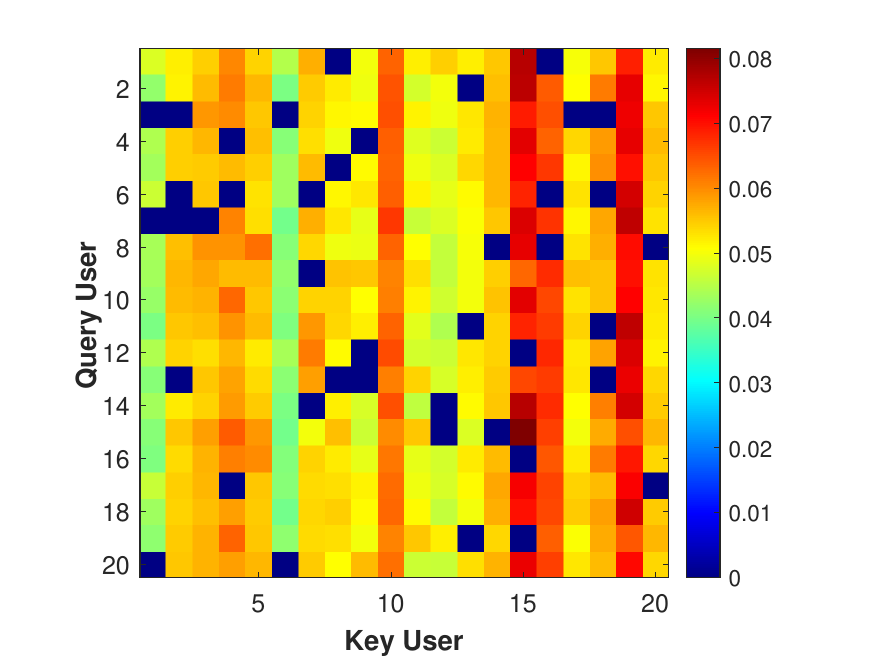}
        \label{ep0}
    }

    \subfloat[Attention weights in the last episode]{%
        \includegraphics[width=\linewidth]{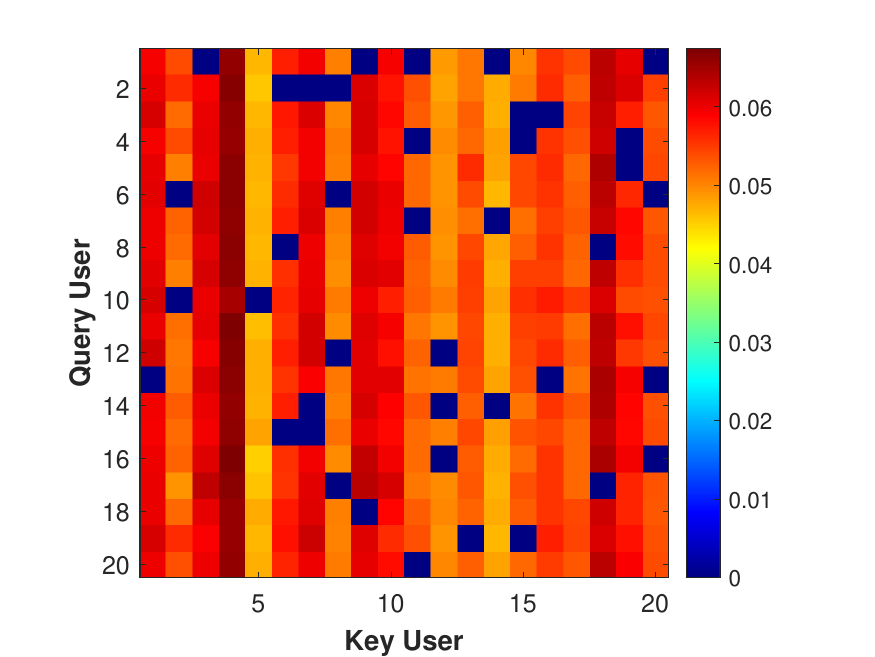}
        \label{ep49000}
    }
    \caption{Attention weights in early and last training.}
    \label{attn_1}
\end{figure}
\subsubsection{Training Dynamics of Attention Weights}

\begin{figure}[htbp]
    \centering

    \subfloat[Attention weights for episode 5000]{%
        \includegraphics[width=\linewidth]{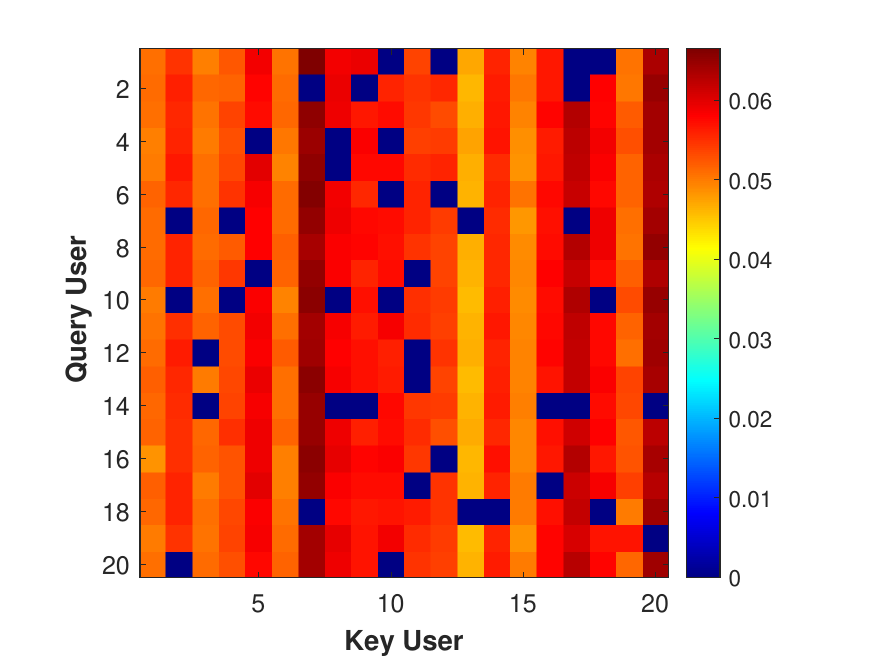}
        \label{ep5000}
    }

    \subfloat[Attention weights for episode 35000]{%
        \includegraphics[width=\linewidth]{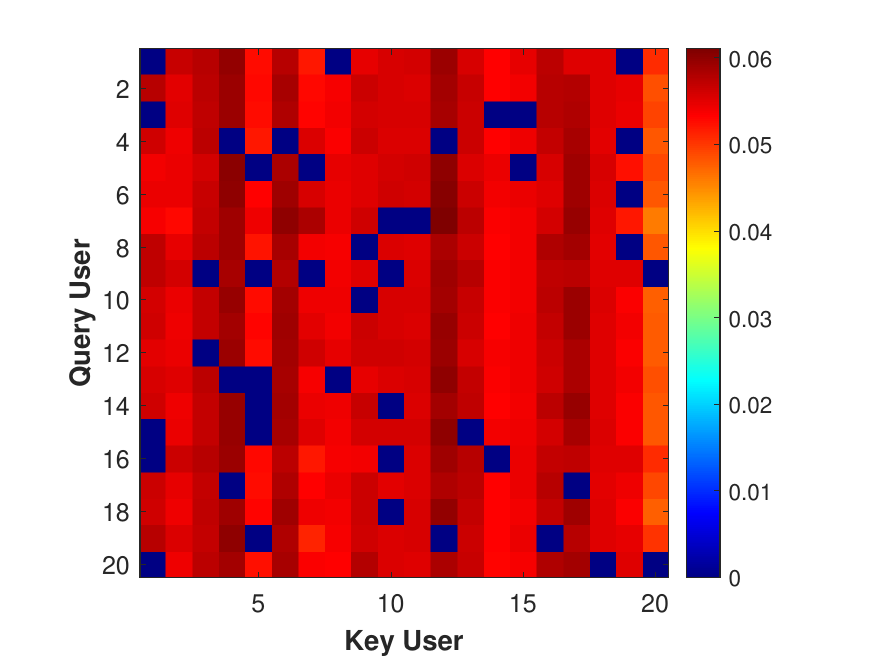}
        \label{ep35000}
    }
    \caption{Attention weights in mid training.}
    \label{attn_2}
\end{figure}

Figures \ref{attn_1} and \ref{attn_2} show attention heatmaps at selected training episodes, where notable trends can be observed:

\begin{itemize}
    \item \textbf{Early Training (Episode 0):} As shown in Fig. \ref{ep0}, the attention weights are broadly distributed with no clear structure. This indicates an untrained policy where all users are treated equally, lacking prioritization or dependency modeling.

    \item \textbf{Mid Training (Episodes 5000–35000):} The attention becomes increasingly sparse and structured, as shown in Fig. \ref{attn_2}. Attention is concentrated on specific users, which suggests that the model is learning to prioritize users based on their impact on the AoI objective. Notably, users with higher priority (e.g., User~1, User~2) receive more attention across rows and columns.

    \item \textbf{Late Training (Episode 50000):} The attention patterns exhibit greater stability with clear focus on particular user pairs or groups (Fig. \ref{ep49000}). This reflects convergence to a learned policy that selectively attends to high-impact users and adapts to the NOMA constraint of at most two users per channel. Some minor redistribution of attention may also occur due to entropy regularization in PPO or fine-tuning of edge-case behaviors.
\end{itemize}

\subsubsection{Interpretation of Attention Patterns}

The learned attention patterns reveal several meaningful insights:

\begin{itemize}
    \item \textbf{Column-wise Attention:} High values in certain columns, especially for User~1 through User~5, indicate that decisions for many users are impacted by the state of  high-priority users. This is desirable as it promotes coordination that respects critical AoI constraints.

    \item \textbf{Row-wise Attention:} High-priority users tend to give more focused attention to others, which suggests they are making more deliberate decisions to maintain low AoI.

    \item \textbf{Emergent User Groupings:} In several episodes, attention appears in localized blocks, potentially indicating learned user pairings under the NOMA constraint. These may reflect effective co-scheduling patterns the agent has discovered.

    \item \textbf{Low Attention to Low-priority Users:} Users with low penalties (e.g., User~20) receive little attention, showing the agent’s ability to ignore users with negligible impact on the reward.
\end{itemize}

The attention mechanism enables the agent to learn a structured, priority-aware scheduling policy that adapts to both user heterogeneity and the NOMA constraint. The emergence of selective attention over time is a strong indicator that the Transformer-based policy is successfully capturing the underlying dependencies between users, leading to more efficient AoI minimization.

\subsection{Computational Complexity} 

While the Transformer encoder improves performance and interpretability, it introduces higher computational complexity than traditional models due to the self-attention mechanism, which scales quadratically with the number of users $U$. The per-layer complexity is $O(U^2 \cdot d_{model})$, which can be challenging for large $U$ in latency-sensitive systems. However, our target scenario involves limited user contention within a NOMA cluster ($U=20$), where the overhead is manageable and the gains are significant. Specifically, the Transformer achieves superior performance by effectively modeling inter-user dependencies, resulting in the lowest AoI and highest reward, and offers interpretability through attention weights that reveal user prioritization and co-scheduling patterns, which is critical for validating policies in URLLC settings.

\section{Conclusion}

This paper proposes a Transformer-based reinforcement learning framework for AoI-aware resource allocation in multi-user NOMA systems. By formulating joint scheduling and power control as an MDP and integrating attention into both the actor and critic, the framework captures asymmetric user importance and inter-user dependencies under heterogeneous task sizes, AoI thresholds, and penalties. Attention map analysis shows that the policy learns to prioritize high-importance users while respecting NOMA constraints, yielding structured and interpretable decision patterns. Overall, the attention-enhanced policy improves learning efficiency, interpretability, and AoI performance, demonstrating the effectiveness of attention-driven coordination in multi-user wireless environments. 

\bibliographystyle{IEEEtran}
\bibliography{IEEEabrv,Bibliography}

\vspace{12pt}

\end{document}